# Secure Iris Authentication Using Visual Cryptography

P.S.Revenkar

Faculty of Department of Computer
Science and Engineering
Government College of Engineering,
Aurangabad, Maharashra, India
prevankar@gmail.com

Anisa Anjum

Department of Computer Science
and Engineering
Government College of Engineering,
Aurangabad, Maharashtra, India
anisa.anjum@gmail.com

W.Z.Gandhare

Principal of Government College of
Engineering,
Aurangabad, Maharashtra , India
wzgandhare@yahoo.com

*Abstract*—**Biometrics deal with automated methods of identifying a person or verifying the identity of a person based on physiological or behavioral characteristics. Visual cryptography is a secret sharing scheme where a secret image is encrypted into the shares which independently disclose no information about the original secret image. As biometric template are stored in the centralized database, due to security threats biometric template may be modified by attacker. If biometric template is altered authorized user will not be allowed to access the resource. To deal this issue visual cryptography schemes can be applied to secure the iris template. Visual cryptography provides great means for helping such security needs as well as extra layer of authentication.**

## I. INTRODUCTION

Biometrics is a technology that uses physiological or behavioral characteristics to authenticate identity of persons [26]. For automated personal identification biometric authentication is getting more attention. There are various application where personal identification is required such as passport control, computer login control, secure electronic banking, bank ATM, credit cards, premises access control, border crossing, airport , mobile phones, health and social services, etc. Many biometric techniques are available such as facial thermogram, hand vein, gait, keystroke, odor, ear, hand geometry, fingerprint, face, retina, iris, palm print, voice and signature. Among those iris recognition is one of the most promising approach because of stability, uniqueness and noninvasiveness [25].

Biometrics systems are more consistent and more user friendly. Still there are certain issues particularly the security facet of both biometric system and biometric data. As template is stored in centralized database, they are vulnerable to eavesdropping and attacks. Thus alternative protection mechanisms need to be considered. For these reasons various researches have been made to protect the biometric data and template in the system by using cryptography, stegnography and watermarking. In this paper a system is proposed by applying visual cryptography technique to biometric template (iris). Visual cryptography technique has been applied on to the iris template to make it secure from attack in centralized database as well as extra layer of authentication to the users.

This paper is organized as follow: Related work for security enhancement of biometrics system and various visual cryptography schemes are discussed in section2, section 3 presents the proposed system, experiments and results are shown in section 4, section 5 provides discussion and section 6 concludes the paper.

## II. RELATED WORK

### A. Security Enhancement Of Biometrics System

Recognizing a person using passwords is not sufficient for reliable identity determination because they can be easily shared, or stolen. A biometric system is essentially a pattern-recognition system that recognizes a person based on a feature vector derived from a specific physiological or behavioral characteristic that the person possesses [26]. Advantages of using biometrics characteristics are reliability, convenience, universality, and so on. But biometrics system does not provide privacy because biometric data is not replaceable and is not secret. There exist several types of attacks possible in a biometric system. Ratha et al.[27] describe eight basic sources of attack as shown in figure 1.

Protecting the template is a challenging task in the biometric system (attack on point 6). Many researches have been made to protect fingerprint and iris data and template [1-10]. Davida at al [1] make the use of error-correcting codes in designing a secure biometrics system for access control. Following the work[1], Juels and Wattenberg [2] broaden the system by establishing a different way of using error-correcting codes and approach is known as "fuzzy commitment". Chander Kant et al[3] presented the idea for biometric security using stegnography to make system more secure . While encoding the secret key (which is in the form of pixel intensities) will be merged in the picture itself, and only the authentic user will be allowed to decode. Khalil Zebbiche et al[4] proposed wavelet-based digital watermarking method to hide biometric data (i.e. fingerprint minutiae data) into fingerprint images. This provides a high security to both hidden data (i.e. fingerprint minutiae) that have to be transmitted and the host image (i.e. fingerprint). To protect fingerprint images K. Zebbiche et al [8] presented an efficient technique for use in fingerprint images watermarking. The underlying principle of the technique is embedding the watermark into the ridges area of the fingerprint images which represents the region of interest. The viability of template-protected biometric authentication systems was exhibited with a fingerprint recognition system by Tuyls et al [9]. A.K.Jain and Uludag.U [10] introduced an amplitude modulation-based watermarking method in which they hide a user's biometric data in a variety of images.

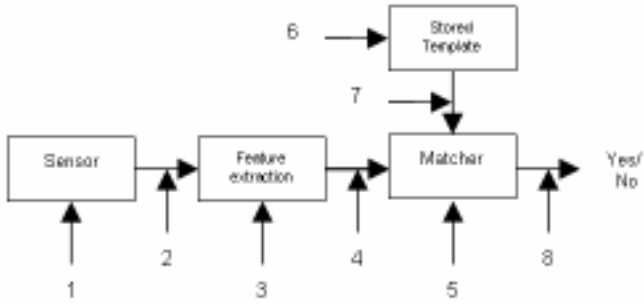

Figure 1.    Possible attack points in generic biometric systems [27]

By combining asymmetric digital watermarking and cryptography as a powerful mechanism was proposed by Nick Bartlow et al [5] to store raw biometric data in centralized databases. Shenglin Yang et al [6] presented a template-protected secure iris verification system based on the Error Correcting Code (ECC) cryptographic technique with the reliable bits selection to improve the verification accuracy. In the scheme a transformed version of the iris template instead of the plain reference is stored for protecting the sensitive biometric data. Jing Dong et al [7] proposed biometric watermarking for protecting biometric data and templates in biometric systems. The scheme suggest protection of iris templates by hiding them in cover images as watermarks (iris watermarks), and protection of iris images by watermarking them.

### B.  Visual Cryptography

The basic visual cryptography scheme was proposed by Naor and Shamir's[11] . In this scheme for sharing a single pixel $p$, in a binary image $Z$ into two shares $A$ and $B$ is illustrated in Table I. If $p$ is white, one of the first two rows of Table 1 is chosen randomly to encode $A$ and $B$. If $p$ is black, one of the last two rows in Table I is chosen randomly to encode $A$ and $B$. Thus, neither $A$ nor $B$ exposes any clue about the binary color of $p$. When these two shares are superimposed together, two black sub-pixels appear if $p$ is black, while one black sub-pixel and one white sub-pixel appear if $p$ is white as indicated in the rightmost column in Table 1. Based upon the contrast between two kinds of reconstructed pixels can tell whether $p$ is black or white. Performance of visual cryptography scheme mainly depends on pixel expansion and contrast. Pixel expansion refers to the number of subpixels in the generated shares that represents a pixel of the original input image. It represents the loss in resolution from the original picture to the shared one. Contrast is the relative difference in weight between combined shares that come from a white pixel and a black pixel in the original image. Plenty of research has been made to improve the performance of basic visual cryptography scheme. Many authors have proposed the visual cryptography schemes in which pixel expansion is 1 [12-18]. These schemes can be used as quality of retrieved images is good.

TABLE I.      ENCODING A BINARY PIXEL P INTO TWO SHARES A AND B USING NAOR AND SHAMIR'S SCHEME

| Z | A | B | A⊗B |
|---|---|---|---|
| | | | |
| | | | |
| | | | |
| | | | |

## III.  PROPOSED SYSTEM

As protecting template in the database securely is one of the challenges in any biometric system. Here visual cryptography technique is applied to iris authentication system. In this system there are two modules: Enrollment module and Authentication module. For accessing any secure resource by authenticated users this system can be used.

### A.  Enrollment

The administrator will collect the eye image of the eligible users those are having access to secure resource. The enrolled eye image is required to be processed so characteristic iris features can be extracted for this purpose algorithms are developed from [21]. Three steps that are: segmentation, normalization, and feature extraction are performed as conferred below:

- Segmentation is performed to extract the iris from the eye image. By employing circular Hough transform boundary of iris is searched. By fitting two lines using the linear Hough Transform eyelids are detected and eyelash is separated by threshold technique.

- Normalization of iris region is carried out using Daugman's rubber sheet model. This model remaps each pixel within the iris region to a pair of polar coordinates. The center of the pupil is considered as the reference point and the radial vectors circle through the iris region.

• Feature extraction is done by convolving the normalized iris pattern into one dimensional Log-Gabor wavelets. The resulting phase information for both the real and the imaginary response is quantized, generating a bitwise template which is of 20*480 size.

In the existing system generated template is stored in the database. As Nalini K. Ratha et al[27] pointed out that the stored template in the database attacker may try to alter result in authorization for a unauthorized users, or denial of service for the authenticated user related with the corrupted template. Here iris template is protected by applying visual cryptography.

For securing iris feature template, the template and another secret binary image which is chosen by system administrator is given as input to the visual cryptography algorithm [15]. Two random shares are created with the help visual cryptography scheme suggested by Wen-Pinn Fang [15]. For sharing two secret images $I_1$ is iris template image (generated from feature template) and $I_2$ is secret image of size 20*480 pixels using the algorithm [15] two shares $S_1$ and $S_2$ are generated. Before starting to generate shares, original image and shares are divided into two same size parts, upper part and lower part: $Image_1^U$, $Image_1^L$, $Image_2^U$, $Image_2^L$, $Share_1^U$, $Share_1^L$, $Share_2^U$ and $Share_2^L$. Following steps are required to generate the shares.

**Step. 1** Assign the pixel values of Share 1U randomly.

**Step. 2** Assign the pixel value of Share 2U.

    if Image 1 [x][y]=white then

    Share 2U [x][y]= Share 1U [x][y].

      else

    Share 2U [x][y]=complement of Share 1U [x][y].

      end if

**Step 3**.Reverse Share 2U,that is

    Temp[x][y]= Share 2U[20-x][y].

**Step 4**. Assign the pixel value of Share 2L.

    if Image 2 [x][y]=white, then

    Share 1L [x][y]= temp[x][y].

      else,

    Share 1L [x][y]=complement of temp[x][y].

      end if

**Step 5.** Assign the pixel value of Share 2L,

    if Image 1[x][y]=white then

    Share 2L [x][y]= Share 1L [x][y].

  else

    Share 2L [x][y]=complement of Share 1L [x][y].

    end if

One share is stored in the database along with user login and other given to user on ID card along with login. Enrollment process is shown in the figure 2. As the visual cryptography techniques guarantee that no information is revealed by one share alone, this provides security to the iris template in the database.

### B. Authentication

For authentication user will provide share in the form of ID card. System finds the corresponding share from database. By stacking two shares first $I_1$ iris template image is created. And from this image iris feature template is generated. The new eye image supplied by user will be processed with three steps: segmentation, normalization and feature extraction which generates iris feature template. Then these two feature templates are matched using hamming distance. If features match access is granted else the verification fails. Authentication process is shown in figure 3.

## IV. EXPERIMENTS AND RESULTS

The most popular and commercial iris recognition system was developed by Daugman [19]. Following this many iris recognition systems are proposed by researchers [20-24]. As main intent of this paper is providing security to the iris template in the database, image processing algorithm for iris feature extraction are derived from [21]. To build this system. MATLAB platform is selected because of powerful inbuilt mathematical, signal and image processing functions for developing algorithm of visual cryptography [15]. Iris images are taken from CASIA Iris Image Database V3.0 [29].

The working of proposed system is shown in figure 4 and 5. For enrollment a single eye image is taken from CASIA database. After performing segmentation, normalization and feature extraction feature template is generated. Iris template image (generated from feature template) and another binary image which is chosen by system administrator is given as input to the visual cryptography algorithm. Two shares are generated Share1 and Share2 as output of visual cryptography

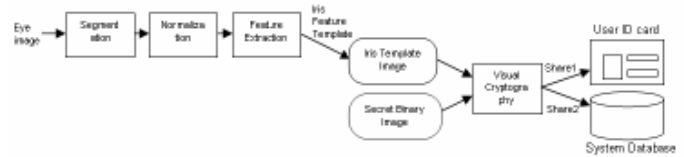

Figure 2.   User Enrollment

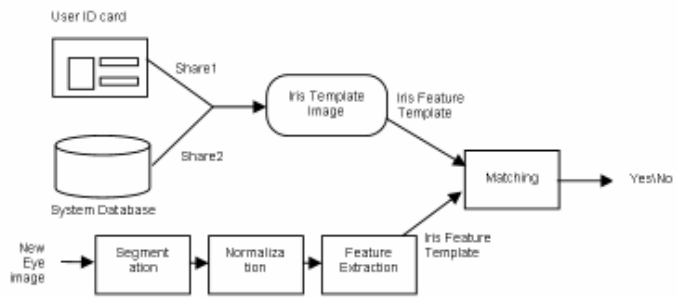

Figure 3.   User Authentication

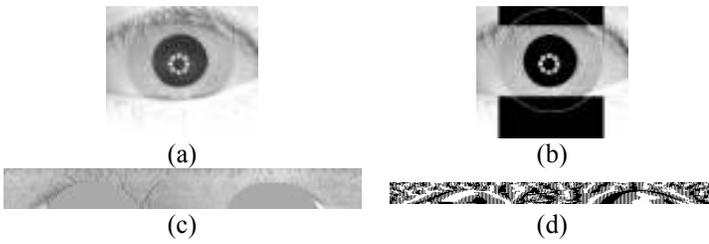

(a)

(b)

(c)

(d)

Figure 4.   (a) Eye Image (b) Iris Segmentation (c) Iris Normalization (d) Extracted feature template

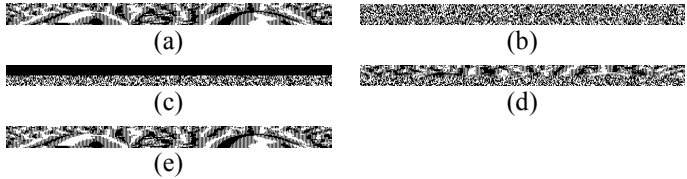

(a)

(b)

(c)

(d)

(e)

Figure 5.   (a) Iris template image (b) Secret Binary Image (c) Share1 (d) Share2 (e) Result of superimposing of share1 and share 2.

algorithm. One share along with username is kept by system and other is given on the user ID Card.

For authentication user provides share which is on the ID card. The share extracted from this card is superimposed with corresponding share that is stored in the database, generates the Iris template image as shown in figure 5 (a-e). From this Iris template image feature template is generated. Now this feature template is matched with Iris feature of newly provided eye image using hamming distance.

## V.   DISCUSSION

The main confront for biometric authentication is to provide a secure storage for the reference template. As Nalini K. Ratha et al [27] described that the stored template in the database may be corrupted by the attacker and resulting in authorization for a unauthorized users, or denial of service for the authenticated user related with the corrupted template. There are various approaches suggested by researchers to store finger print and iris data and template securely using various techniques like cryptography, stegnography and watermarking [1-10]. Previously Y.V. Subba Rao et al [28] has applied the visual cryptography techniques to the area of authentication using fingerprints.

In the proposed system visual cryptography techniques is applied to protect iris template in the database as well as providing extra layer of authentication to the existing iris authentication system. As enrolled iris template is divided into two shares using visual cryptography one is kept in the database and other with user on the ID card. Security is provided to the iris template because using the only one share which is in the database no information can be retrieved for the enrolled eye image. In this case access from unauthorized user is avoided. This system will be more secure and reliable in security-critical applications.

## VI.   CONCLUSION AND FUTURE WORK

Various approaches adopted by researchers to secure the raw biometric data and template in database are discussed here. In this paper a method is proposed to store iris template securely in the database using visual cryptography. Experimental results indicate that by applying visual cryptography techniques on iris template for more security, matching performance of iris recognition is unaffected with extra layer of authentication. Speed of iris authentication system is slower [21] it can be also improved using other systems. Here generated shares are meaningless using other visual cryptography techniques which generates meaningful share can also be applied.